\documentstyle[psfig,sprocl]{article}

\def\be{\begin{equation}}
\def\ee{\end{equation}}
\def\bea{\begin{eqnarray}}
\def\eea{\end{eqnarray}}
\begin{document}

\title{QUANTUM FEEDBACK FOR PROTECTION OF SCHR\"ODINGER CAT STATES}

\author{M. FORTUNATO, P. TOMBESI, D. VITALI}

\address{Dipartimento di Matematica e Fisica, Universit\`a di
Camerino, via Madonna delle Carceri I-62032 Camerino \\
and INFM, Unit\`a di Camerino, Italy \\ E-mail: mauro@camcat.unicam.it} 

\author{J. M. RAIMOND}

\address{Laboratoire Kastler Brossel, D\'epartement de Physique
de l'Ecole Normale Sup\'erieure, \\ 
24 rue Lhomond, F-75231 Paris Cedex 05, France}

%%%%%%%%%%%%%%%%%%%%%%%%%%%%%%%%%%%%%%%%%%%%%%%%%%%%%%%%%%%%%%
% You may repeat \author \address as often as necessary      %
%%%%%%%%%%%%%%%%%%%%%%%%%%%%%%%%%%%%%%%%%%%%%%%%%%%%%%%%%%%%%%

\maketitle\abstracts{We review the use of quantum feedback for 
combatting the decoherence of Schr\"odinger-cat-like states in
electromagnetic cavities, with special emphasys on our recent
proposal of an automatic mechanism based on the injection of
appropriately prepared ``probe'' and ``feedback'' Rydberg atoms.
In the latter scheme, the information transmission from the
probe to the feedback atom is directly mediated by a second
auxiliary cavity. The detection efficiency for the probe atom
is no longer a critical parameter, and the decoherence time of
the linear superposition state can be significantly increased
using presently available technology.}

\section{Introduction}

One of the most fundamental issues in quantum theory is how the
classical macroscopic world emerges from the quantum substrate.
This question is also an important point in the interpretation of
quantum mechanics and it is still the subject of an intense
debate~\cite{zur,ZEH98}. The most striking example of this problem
is given by the possibility, opened by quantum mechanics, of having
linear superpositions of macroscopically distinguishable states, the 
so-called ``Schr\"odinger-cat'' states. 
Such paradoxical states are very sensitive to {\em decoherence}, i.e.,
the rapid transformation of these linear superpositions into the
corresponding classical statistical mixture, caused by the unavoidable
entanglement of the system with uncontrolled degrees of freedom of the
environment~\cite{zur}. The decoherence time depends on the form of
system-environment interaction~\cite{anglin} but, in most cases, it is
inversely proportional to the squared ``distance'' between the two
states of the superposition~\cite{milwal}.
It is then clear, that for macroscopically distinguishable states, the 
decoherence process becomes thus practically instantaneous~\cite{zur}. 
Decoherence is therefore experimentally accessible only in the {\em 
mesoscopic} domain. In this case, one is able to monitor the progressive
emergence of classical properties from the quantum ones. In this 
context, an important achievement has been obtained by Monroe {\it et
al.}~\cite{wine}, who have prepared a trapped ${\rm ^{9}Be^{+}}$ ion
in a superposition of spatially separated coherent states and
detected the quantum coherence between the two localized states.
However, the decoherence of the superposition state has not been 
observed in this experiment. 
The progressive decoherence of a mesoscopic Schr\"odinger cat has been 
monitored for the first time in the experiment of Brune
{\it et al}.~\cite{prlha}, where the linear superposition of two coherent
states of the electromagnetic field in a cavity with classically distinct
phases has been generated and detected.

Recently, the field of quantum information theory~\cite{bennet} has 
undergone an impressive development, and the study of decoherence has 
then become important not only from a fundamental, but also from a more
practical point of view. All the quantum information processing applications
rely on the possibility of performing unitary transformations on a system of
$N$ quantum bits (qubits), whose decoherence has to be made as small as possible. As a
consequence, decoherence control is now a rapidly expanding field of
investigation. In this respect, quantum error correction codes~\cite{error}
have been developed in which the entangled superposition state of $N$ qubits is 
``encoded'' in a larger number of qubits. Assuming that only a fraction of
qubits decoheres, it is then possible to reconstruct the original state with a
suitable decoding procedure.
These codes always require the entanglement of a large number of qubits, and
will become practical only if quantum networks of tens of qubits become
available. Up to now, the polarization states of three photons have been
entangled at most~\cite{ghz}. Entangled states of two Rydberg
atoms~\cite{EPRPAIR} or of two trapped ions~\cite{turch} at most can be
generated.
Therefore, in the present experimental situation, it is more realistic to
study complementary and more ``physical'' ways to deal with decoherence, based
on the explicit knowledge of the specific process causing decoherence, which
could be applied with very few degrees of freedom. This is possible, in
particular, in quantum optics, when information is encoded in the quantum
states of an electromagnetic mode (see for example~\cite{milb}). In this case 
decoherence is caused by photon leakage.
It is then possible to develop experimental schemes able to face photon leakage 
and the associated decoherence.

We have already shown in some recent papers~\cite{homo,minsk,prlno,jmo,pran}
that a possible way to control decoherence in optical cavities is given by
appropriately designed feedback schemes. Refs.~\cite{homo} show that a
feedback scheme based on the continuous homodyne measurement of an optical
cavity mode is able to increase the decoherence time of a superposition state.
In Ref.~\cite{jmo,pran} a feedback scheme based on continuous photodetection
and the injection of appropriately prepared atoms has been considered. This
scheme, in the limit of very good detection efficiency, is able to obtain a
significant ``protection'' of a generic quantum state in a cavity.
In~\cite{prlno,pran} this photodetection-mediated scheme has been adapted to
the microwave experiment of Ref.~\cite{prlha} in which photodetectors cannot
be used. The cavity state can only be indirectly inferred from measurements
performed on probe atoms which have interacted with the cavity mode.
Under ideal conditions, this adaptation to the microwave cavity case 
leads to a significant increase of the lifetime of the Schr\"odinger cat
generated in~\cite{prlha}. However, this scheme suffers from two important
limitations, making it very inefficient when applied under the actual
experimental situation. 
It first requires the preparation of samples containing {\em exactly} one
Rydberg atom sent through the apparatus. Up to now, the experimental 
techniques allow only to prepare a sample containing a random atom number,
with a Poisson statistics. Two-atom events are excluded only at the expense
of a low average atom number, lengthening the feedback loop
cycletime~\cite{EPRPAIR}. The original scheme requires also a near unity
atomic detection efficiency, which is extremely difficult to achieve even
with the foreseeable improvements of the experimental apparatus.

Here we propose a significant improvement of the microwave feedback scheme
described in \cite{prlno,pran}. This new version, using a direct transmission
of the quantum information from the probe to the feedback atom, does not
require a large detection efficiency, removing one of the main difficulties
of the previous design. It however also requires sub-poissonian atom statistics. 
We show briefly how such atomic packets could be in principle prepared with
standard laser techniques. Finally, our scheme improves the efficiency of the 
feedback photon injection in the cavity by using an adiabatic rapid passage.

\section{Detection-mediated feedback}

In this section we briefly review the original ``stroboscopic'' feedback
scheme for microwave cavities proposed in~\cite{prlno,pran}. 
This proposal is based on a very simple idea: whenever the cavity looses a 
photon, a feedback loop supplies the cavity mode with another photon, through
the injection of an appropriately prepared atom.
However, since there are no good enough photodetectors for microwaves, 
one has to find an indirect way to check if the high-Q microwave cavity has
lost a photon or not. In the experiment of Brune {\it et al.}~\cite{prlha},
information on the cavity field state is obtained by detecting the state of a
circular Rydberg atom which has dispersively interacted with the
superconducting microwave cavity. This provides an ``instantaneous"
measurement of the cavity field and suggests that continuous photodetection
can be replaced by a series of {\it repeated} measurements, performed by 
non-resonant atoms regularly crossing the high-Q cavity.

The experimental scheme of the stroboscopic feedback loop is a
simple modification of the scheme employed in Ref.~\cite{prlha}.
The relevant levels of the velocity-selected atoms are two adjacent circular
Rydberg states with principal quantum numbers $n=50$ and $n=51$ (denoted by 
$|g\rangle$ and $|e\rangle$, respectively) and a very long lifetime 
($\simeq 30$ ms). 
The high-Q superconducting cavity is sandwiched between two low-Q 
cavities $R_{1}$ and $R_{2}$, in which classical microwave fields resonant 
with the transition between $|e\rangle$ and $|g\rangle$ can be applied. 

The high-Q cavity $C$ is instead slightly off-resonance 
with respect to the $e\, \rightarrow \, g$ transition, with a detuning
$\delta = \omega - \omega_{eg}$, where $\omega$ is the cavity mode frequency
and $\omega_{eg}=(E_{e}-E_{g})/\hbar $. The Hamiltonian of the 
atom-microwave cavity mode system is the Jaynes-Cummings Hamiltonian,
\begin{equation}
H_{JC}=E_{e}|e\rangle \langle e | + E_{g}|g\rangle \langle g |+
\hbar \omega a^{\dagger} a
+\hbar \Omega \left(|e \rangle \langle g |a+|g \rangle \langle e |
a^{\dagger}\right) \;,
\label{jc}
\end{equation}
where $\Omega $ is the vacuum Rabi coupling between the atomic dipole 
on the $e\, \rightarrow \, g$ transition and the cavity mode.
In the off-resonant case and perturbative limit $\Omega \ll \delta $, the 
Hamiltonian~(\ref{jc}) assumes the dispersive form \cite{pran,harray,brune}
\begin{equation}
H_{disp}=\hbar \frac{\Omega ^{2}}{\delta} \left(|g\rangle
\langle g| a^{\dagger} a-|e \rangle \langle e | 
a^{\dagger} a \right)\;.
\label{heff}
\end{equation}

A linear superposition state of two coherent states with opposite phases is
generated when the cavity mode is initially in a coherent state
$|\alpha \rangle $ and the Rydberg atom, which is initially prepared in the
excited level $|e\rangle $, is subjected to a $\pi/2$ pulse both in $R_{1}$
and in $R_{2}$. In fact, when the atom has left the cavity $R_{2}$, the joint
state of the atom-cavity system becomes the entangled
state~\cite{prlha,pran,brune} 
\begin{equation}
|\psi_{atom+field}\rangle = \frac{1}{\sqrt{2}}\left(|e\rangle 
\left(|\alpha e^{i\phi}\rangle -|\alpha e^{-i\phi}\right)
+|g\rangle \left(|\alpha e^{i\phi}\rangle +|\alpha e^{-i\phi}\right) \right)\;,
\label{atf1}
\end{equation}
where $\phi = \Omega ^{2}t_{int}/\delta $ and $t_{int}$ is the
interaction time in $C$.
A Schr\"odinger-cat state is then conditionally generated in the microwave
cavity as soon as one of the two circular atomic states is detected. 

As it was shown in Ref.~\cite{pran}, the stroboscopic feedback scheme
works only for Schr\"odinger cat states with a definite parity, i.e.
even or odd cat states, and therefore we shall restrict to $\phi = \pi/2$
from now on. In fact, when the cavity field initial state is a generic
density matrix $\rho$, the state of the probe atom-field system after
the two $\pi/2$ pulses and the $\phi=\pi/2$ conditional phase-shift can be
written as \cite{pran}
\begin{equation}
\rho_{atom+field}= |e\rangle \langle e| \otimes \rho_{e} +
|g\rangle \langle g| \otimes \rho_{g} 
+|e\rangle \langle g| \otimes \rho_{+} +
|g\rangle \langle e| \otimes \rho_{-} \;,
\label{2at3}
\end{equation}
where 
\begin{eqnarray}
\rho_{e} &=& P_{odd}\rho P_{odd}
\label{odd} \\
\rho_{g} &=& P_{even}\rho P_{even} \;,
\label{even} 
\end{eqnarray}
are the projections of the cavity field state onto the subspace with an odd
and even number of photons, respectively, and the operators $\rho_{\pm}$
(whose expression is not relevant here) are given in~\cite{pran}.
Eq.~(\ref{2at3}) shows that there is a perfect correlation between the atomic
state and the cavity field parity, which is the first step in an optimal 
quantum non demolition measurement of the photon number~\cite{OPTQND}.
It is possible to prove that this perfect correlation between the atomic
state and a cavity mode property holds only in the case of an exact 
$\phi=\pi/2$-phase shift sandwiched by two classical $\pi/2$ pulses in cavities
$R_{1}$ and $R_{2}$~\cite{pran}.
Moreover, the entangled state of Eq.~(\ref{2at3}) allows to understand how it
is possible to check if the microwave cavity $C$ has lost a photon or not and
therefore to trigger the feedback loop, using atomic state detection only.
The detection of $e$ or $g$ determines the parity of the field and, provided
that the probe atomic pulses are frequent enough, indicates whether a
microwave photon has left $C$ or not.
In fact, let us consider for example the case in which an odd cat state is
generated (first atom detected in $e$): a probe atom detected in state $e$
means that the cavity field has remained in the odd subspace. The cavity has
therefore lost an {\it even} number of photons. If the time interval
$\tau_{pr}$ between the two atomic pulses is much smaller than the cavity
decay time $\gamma^{-1}$, $\gamma \tau_{pr} \ll 1$, the probability of
loosing two or more photons is negligible and this detection of the probe
atom in $e$ means that no photon has leaked out from the high-Q cavity $C$.
On the contrary, when the probe atom is detected in $g$, the cavity mode
state is projected into the even subspace. The cavity has then lost an
{\it odd} number of photons. Again, in the limit of enough closely 
spaced sequence of probe atoms, $\gamma \tau_{pr} \ll 1$, the probability of 
loosing three or more photons is negligible. A detection in $g$ means that
one photon has exited the cavity. Therefore, for achieving a good protection
of the initial odd cat state, the feedback loop has to supply the
superconducting cavity with a photon whenever the probe atom is detected in
$g$, while feedback must not act when the atom is detected in the $e$ state.

In Ref.~\cite{pran} it has been proposed to realize this feedback loop with
a switch connecting the $g$ state field-ionization detector with a second
atomic injector, sending an atom in the excited state $e$ into the high-Q
cavity. The feedback atom is put in resonance with the cavity mode by another 
switch turning on an electric field in the cavity $C$ when the atom enters it, 
so that the level $e$ is Stark-shifted into resonance with the cavity mode.

As it is shown in Ref.~\cite{pran}, if the probe atomic pulses are
sufficiently frequent, this stroboscopic feedback scheme becomes extremely
efficient and one gets a good preservation of an initial Schr\"odinger-cat
state. However, if we consider the adaptation of this scheme to the present
experimental apparatus of Ref.~\cite{prlha}, we see that it suffers from two
main limitations, which significantly decrease its efficiency. First of all
the scheme is limited by the non-unit efficiency of the atomic state detectors 
($\eta_{det}\simeq 0.4$), since the feedback loop is triggered only when the
$g$-detector clicks. Most importantly, the above scheme assumes one has
perfect ``atomic guns'', i.e. the possibility of having probe and feedback
atomic pulses with {\it exactly one atom}. This is not experimentally
achieved up to now. The actual experiment \cite{prlha} has been performed
using atomic pulses with a probability of having exactly one atom
$p_{1} \simeq 0.2$, close to the mean atom number in the sample.
This low mean atom number has been chosen to minimize two--atom events. In
this experimental situation, the proposed stroboscopic feedback scheme would 
have an effective efficiency $\eta_{eff}=\eta_{det}p_{1}^{2} \simeq 0.016$, 
too low to get an appreciable protection of the Schr\"odinger cat state.
In the next section we show how this scheme may be improved and adapted to
the experimental apparatus employed in Ref.~\cite{prlha}.

\section{The new automatic feedback scheme}

\begin{figure}[ht]
\centerline{\psfig{figure=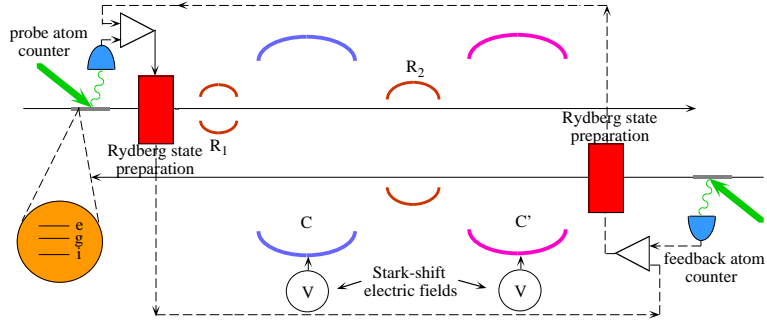,width=4in,angle=270}}
\caption{Schematic diagram of the autofeedback scheme
proposed in this paper. $R_{1}$ and $R_{2}$ are the two 
cavities in which classical microwave pulses can be applied, $C$ is 
the microwave cavity of interest and $C'$ is the cavity automatically 
performing the needed correction. Electric fields can be applied at 
the superconducting mirrors of $C$ and $C'$ to Stark shift the 
Rydberg levels in order to tune the interaction times in $C'$ and 
realize adiabatic transfer in $C$.}
\label{appa}
\end{figure}

From the above discussion, it is clear that the limitations due to the
non-unit efficiency of the atomic detectors could be avoided if we eliminate
the measurement step in the feedback loop and replace it with an
``automatized'' mechanism preparing the correct feedback atom whenever 
needed. This mechanism can be provided by an appropriate
conditional quantum dynamics, and this, in turn, may be provided by a
second high-Q microwave cavity $C'$, similar to $C$, replacing the atomic
detectors, crossed by the probe atom first and by the feedback atom soon
later, as described in Fig.~\ref{appa}.

The cavity $C'$ is resonant with the transition between an auxiliary circular
state $i$ (an immediately lower circular Rydberg state), and level $g$.
The interaction times have to be set so that both the probe and the 
feedback atom experience a $\pi$ pulse when they cross the empty cavity $C'$
in state $g$ (or when they enter in state $i$ with one photon in $C'$). 
This interaction copies the state of the probe atom onto the feedback atom,
and thus removes any need for a unit detection efficiency.

This fine tuning of the interaction times to achieve the $\pi$-spontaneous
emission pulse condition can be obtained applying through the
superconducting mirrors of $C'$ appropriately shaped Stark-shift electric
fields which put the atoms in resonance with the cavity mode in $C'$ only for
the desired time. In this way, since $C'$ is initially in the vacuum state,
when the probe atom crosses $C'$ one has
\begin{eqnarray}
|e\rangle _{p} |0\rangle _{C'} &\rightarrow & |e\rangle _{p} |0\rangle _{C'}\;,
\label{prob1} \\
|g\rangle _{p} |0\rangle _{C'} &\rightarrow & |i\rangle _{p} |1\rangle _{C'}\;.
\label{prob2}
\end{eqnarray}
Soon later a feedback atom enters $C'$ in the state $|i\rangle_{f}$ and one has
\begin{eqnarray}
|i\rangle _{f} |0\rangle _{C'} &\rightarrow & |i\rangle _{f} |0\rangle _{C'}\;,
\label{fb1} \\
|i\rangle _{f} |1\rangle _{C'} &\rightarrow & |g\rangle _{f} |0\rangle _{C'}\;.
\label{fb2}
\end{eqnarray}
In this way the cavity $C'$ is always left disentangled in the vacuum state. 
The feedback atom exiting $C'$ in $|g\rangle $ can be promoted to 
$|e\rangle $ before entering $C$, as required by the feedback scheme, by
subjecting it to a $\pi$ pulse in the classical cavity $R_{2}$
(see Fig.~\ref{appa}).
Therefore, the conditional dynamics provided by $C'$ eliminates any limitation
associated to the measurement and leads to an ``automatic feedback'' 
scheme with unit efficiency in principle.
Also the second important limitation of the stroboscopic feedback scheme
of~\cite{prlno,pran}---i.e. that it requires exactly one probe and one
feedback atom per loop---can be circumvented: A better control of the atom
number, providing single atom events with a high probability, could be achieved
by a modification of the Rydberg atoms preparation technique, in such 
a way that it is only triggered when the fluorescence detection signal
(see Fig.~\ref{appa}) provides evidence of only one atom in the beam,
implementing an atom counter.

Instead of preparing a random atom number at a given time, one thus
prepares with a high probability a single Rydberg atom after a random delay.
However, after a full quantum mechanical calculation and lengthy 
algebra, it is possible to determine the map of a generic feedback cycle,
that is, the transformation connecting the states of the cavity field in $C$
soon after the passage of two successive feedback atoms in $C$, which 
also takes into account the non-unit efficiency of the Rydberg state
preparation. This map, which is reported elsewhere~\cite{noi}, allows 
us to study the dynamics of the Schr\"odinger-cat state in the 
presence of feedback, and to compare it with the corresponding 
dynamics in absence of feedback.

\begin{figure}[ht]
\centerline{\psfig{figure=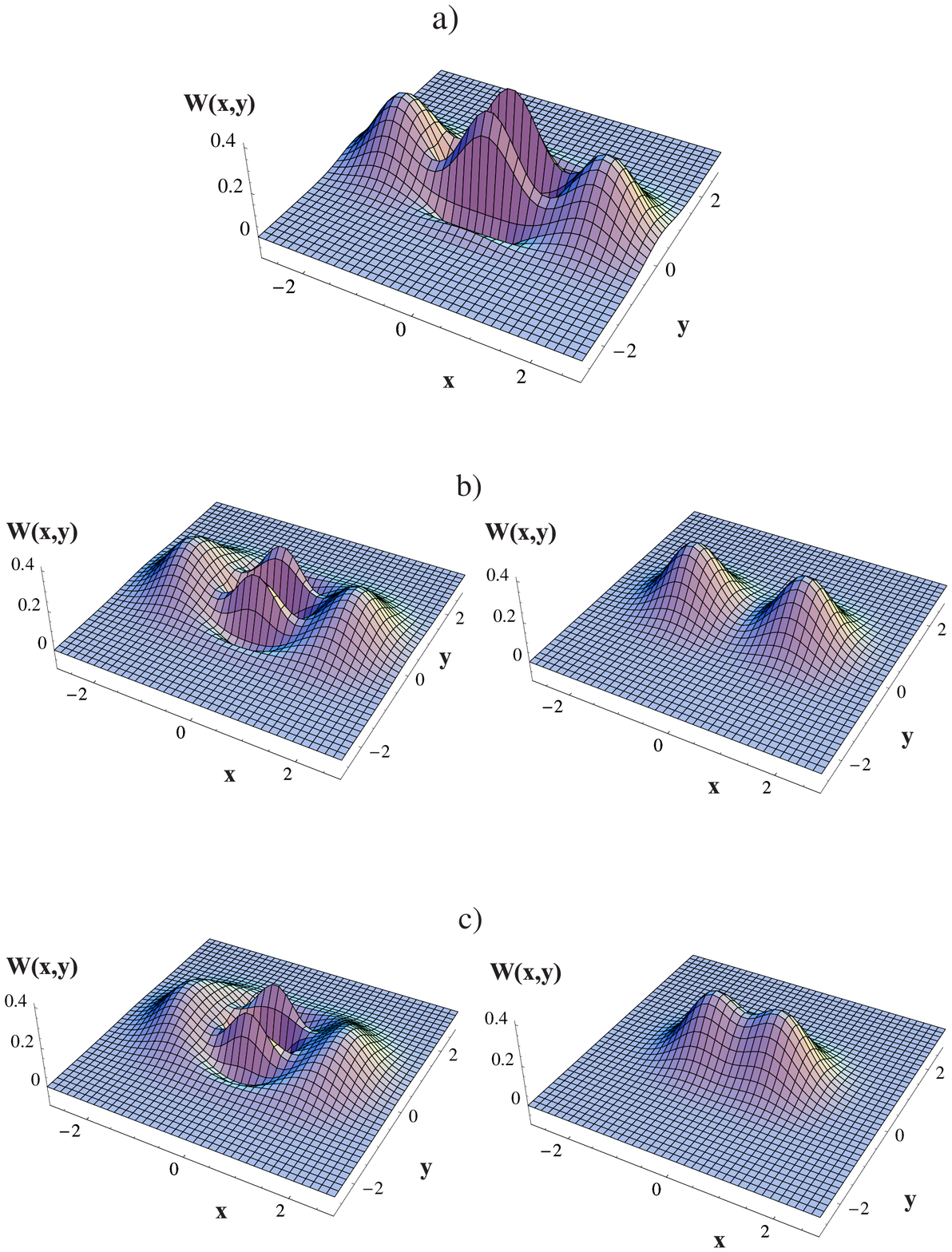,width=3.8in}}
\caption{Wigner function of the initial odd cat state,
         $|\psi \rangle = N_{-}(|\alpha \rangle - |-\alpha \rangle )$,
         with $|\alpha |^{2}=3.3$ (a, top).
         Wigner function of the same cat
         state after $13$ feedback cycles (b), corresponding to a mean
         elapsed time $\bar{t} \simeq 1/\gamma \simeq 6.6 t_{\rm dec}$,
         and after $25$ feedback cycles (c) corresponding to a mean elapsed time
         $\bar{t} \simeq 2/\gamma \simeq 13 t_{\rm dec}$ (left).
         Wigner function of the same cat state after one relaxation time
         $t=1/\gamma$ (b), and after two relaxation times  
         $t=2/\gamma$ (c), in absence of feedback (right).}
\label{strobo}
\end{figure}

In Fig.~\ref{strobo} we show the Wigner function of the initial odd
cat state (top) and its dynamics in presence (left) and in absence
(right) of feedback. The comparison between the two performances is
striking: in absence of feedback the Wigner function becomes quickly
positive definite, while in the presence of feedback the quantum
aspects of the state remain well visible for many decoherence times.

\section{Conclusions}
\label{conclu}

In this paper we have proposed a method to significantly increase the
``lifetime'' of a Schr\"odinger cat state of a microwave cavity mode.
However, as it can be easily expected, most of
the techniques presented here could be applied to the case of a generic
quantum state of a cavity mode (see also Ref.\cite{pran}).
After the first experimental evidences of decoherence mechanisms, 
decoherence control is an expanding field in quantum physics.
An experimental realization of this realistic feedback scheme
would be an important step in this direction.

\section*{Acknowledgments}

This work has been partially supported by INFM (through 
the 1997 Advanced Research Project ``CAT''), by the
European Union in the framework of the TMR Network ``Microlasers
and Cavity QED'' and by MURST under the ``Cofinanziamento 1997''.

\end{document}